\documentclass{article}
\usepackage{eurosym}
\usepackage{amssymb}
\usepackage{amsmath}
\usepackage{cite}

\input{tcilatex}

\begin{document}

\title{Singularity analysis and analytic solutions for the Benney-Gjevik
equations}
\author{Andronikos Paliathanasis\thanks{%
Email: anpaliat@phys.uoa.gr} \\
{\ \textit{Institute of Systems Science, Durban University of Technology }}\\
{\ \textit{PO Box 1334, Durban 4000, Republic of South Africa}} \\
\and Genly Leon \\
{\ \textit{Departamento de Matem\'{a}ticas, Universidad Cat\'{o}lica del
Norte, Avda. }}\\
{\ \textit{A}}\textit{ngamos 0610, Casilla 1280 Antofagasta, Chile } \and %
P.G.L. Leach \\
{\ \textit{Institute of Systems Science, Durban University of Technology }}\\
{\ \textit{PO Box 1334, Durban 4000, Republic of South Africa}} \\
{\ \textit{School of Mathematical Sciences, University of KwaZulu-Natal }}\\
{\ \textit{Durban, Republic of South Africa}}}
\maketitle

\begin{abstract}
We apply the Painlev\'{e} Test for the Benney and the Benney-Gjevik
equations which describe waves in falling liquids. We prove that these two
nonlinear $1+1$ evolution equations pass the singularity test for the
travelling-wave solutions. The algebraic solutions in terms of Laurent
expansions are presented.

\end{abstract}

\section{Introduction}

In 1966 in a short article Benney \cite{benney66} proved the existence of
two-dimensional waves with finite amplitude in the case of a laminar flow of
a thin liquid film down an inclined plane by studying the nonlinearised
system. The total set of differential equations which describe the problem
is the equation of motion for the laminar flow and the continuity equation
for the liquid film \cite{benney66,ben1,ben2,mei1,book1,book2}.

By using the approach described in \cite{ben1} and considering nonlinear
terms Benney \cite{benney66} reduced the problem to a third-order nonlinear $%
1+1$ evolution equation \cite{benney66}, up to a second-order correction,
namely 
\begin{equation}
u_{t}+A\left( u\right) u_{x}+\varepsilon \left( B\left( u\right)
u_{xx}+C\left( u\right) u_{x}^{2}\right) +\varepsilon ^{2}\left( D\left(
u\right) u_{xxx}+E\left( u\right) u_{x}u_{xx}+F\left( u\right)
u_{x}^{3}\right) =0.  \label{eq.01}
\end{equation}%
Functions $A\left( u\right)$, $B\left( u\right) ,...$ are presented in
Appendix A, which depend also upon the Rayleigh number, $R$. The parameter $%
\varepsilon $ is taken to be an infinitesimal, i.e. $\varepsilon
^{2}\rightarrow 0$, but for our analysis we consider it as a normal
parameter. Finally the dependent variable ,$u\left( t,x\right), $ describes
the location of the free surface of the liquid film. For a review on the
Benney equation see \cite{rev1,rev2}.

While the analysis of Benney \cite{benney66} includes the linear case of 
\cite{ben1,ben2}, the effect of the surface tension on the whole evolution
of the problem has been omitted. However, such an analysis was later
performed in 1970 by Gjevik \cite{gjevik}.

In a similar way as Benney, Gjevik was able to reduce the problem to the
nonlinear $1+1$ evolution equation%
\begin{equation}
u_{t}+A\left( u\right) u_{x}+\varepsilon \left( B\left( u\right)
u_{xx}+C\left( u\right) u_{x}^{2}\right) +\varepsilon ^{2}\left( \bar{D}%
\left( u\right) u_{x}u_{xxx}+\bar{E}\left( u\right) u_{xxxx}\right) =0.
\label{eq.02}
\end{equation}%
In the linear consideration, $\varepsilon ^{2}\rightarrow 0,$ equations (\ref%
{eq.01}) and (\ref{eq.02}) are identical However, the main difference
between these two evolution equations is that (\ref{eq.01}) is of
third-order while (\ref{eq.02}) is of fourth-order\footnote{%
Any differential equation on falling liquid films which is derived with the
approach described in \cite{benney66} is termed a Benny equation. However,
on his original paper Benney provides a third-order derivative in the first
nonlinear correction $\varepsilon ^{2}$.}.

The approach that Benney followed in order to study the existence of wave
solutions for equation (\ref{eq.01}) is based on an approximate solution
around a stationary point, $u_{0}$, and more specifically he considered $%
u\left( t,x\right) =1+\omega \eta \left( t,x\right) $ and linearised around $%
\omega $. However, such an analysis fails to provide an answer to whether
the Benney equation is integrable. In this work, we study the integrability
of the Benney-Gjevik equations (\ref{eq.01}) and (\ref{eq.02}) by using the
method of singularity analysis for differential equations \cite{ramani1}.

Singularity analysis is a powerful tool to study the integrability of
differential equations and has a wide range of applications in mathematical
physics, for instance see \cite{sing1,sing2,sing3,sing4} and references
therein. As far as concerns $1+1$ evolution equations, Gl\"{o}ckle et al in 
\cite{gloglo} present a detailed application on the Painlev\'{e} Test for a
wide range of nonlinear equations and they show how the results of the
Painlev\'{e} Test can be used to find new similarity solutions. A similar
analysis was done by Steeb et al in \cite{sing5} a few years earlier. For
other applications of the singularity analysis on diffusion equations we
refer the reader to \cite{sing6,sing7,sing8}. The plan of the paper follows.

In \ref{sec2} we present the basic properties and definitions for the
singularity analysis while a demonstrative example is given. Sections \ref%
{sec3} and \ref{sec4} include the main results of our analysis in which we
prove the integrability of the Benney and the Benney-Gjevik equations and we
present the algebraic solution. Finally, we draw our conclusions in Section %
\ref{sec5}.

\section{Preliminaries}

\label{sec2}

The development of the Painlev\'{e} Test for the determination of
integrability of a given equation or system of equations and its
systematization has been succinctly summarized by Ablowitz, Ramani and Segur
in the so-called ARS algorithm \cite{ars1,ars2,ars3}. The main steps of the
ARS algorithm are described as: (a) first step is to find the leading-order
behaviour, (b) second step is the derivation of the resonances and (c) final
step is the consistency test. The determination of the leading-order term
provides the movable pole for the differential equation, while the
resonances describe the position of the constants of integration and also
define the explicit form of the Laurent expansion. Finally the consistency
test proves that the Laurent expansion solves the original differential
equations. More details on the method and various discussions can be found
in the review of \cite{ramani1}, while a recent discussion on the connection
of the singularity analysis with the symmetry approach is given in \cite{le1}%
.

The ARS algorithm is somewhat like a recipe which can be difficult to
implement at times. These difficulties can be due to the inherent complexity
of the equation under consideration or it can be due to results at variance
with the prevailing canon for an equation which is patently integrable. This
has lead to changes in the paradigms of singularity analysis. Originally the
singularity had to be a pole, where the alternate description as the method
of polelike expansions. Singularity analysis is valid when the leading-order
term has as exponent power a fraction number or a negative integer number.
On the other hand, for a positive fractional power by inverse the dependent
variable it eventually leads to a negative exponent and hence into a pole.
As the analysis is undertaken over the complex plane of the independent
variable, the presence of fractional exponents means the division of the
plane by branch cuts. A practical limitation would be not to have a large
denominator thereby leading to many subsets of the complex plane. Another
fallacy was that resonances, apart from the generic $-1$, had to be
nonnegative. This was demonstrated to be not so with some elementary
equations the closed-form solutions of which could be used to valid the
point. Finally the occurrence of multiple $-1$ resonances was taken to mean
that the solution had to had a logarithmic singularity. This was shown by
explicit example not to be so.

It is straightforward to find that the application of Lie's theory \cite%
{bluman} for the two $1+1$ PDEs of our consideration provides the two
symmetry vectors $\Gamma _{1}=\partial _{t}$ and $\Gamma _{2}=\partial _{x}$
for both equations. The application of the invariants obtained by these
symmetries reduces the PDEs to ODEs. Static solutions follow by applying the
symmetry vector $\Gamma _{1}$ and stationary solutions follow by applying
the invariants of the symmetry vector $\Gamma _{2}$. In our case, as we are
interested in travelling-wave solutions we apply the invariant functions of
the the symmetry vector $\Gamma _{1}+c\Gamma _{2}$ which are $\zeta =x-ct,~$ 
$y=u\left( \zeta \right) $, in which $c$ is the wave speed.

By assuming $\zeta $ to be the new independent variable and $u\left( \zeta
\right) $ the new dependent variable, the Benney equation reduces to the
third-order ODE%
\begin{equation}
\left( A\left( y\right) -1\right) y^{\prime }+\varepsilon \left( B\left(
y\right) y^{\prime \prime }+C\left( y\right) \left( y^{\prime }\right)
^{2}\right) +\varepsilon ^{2}\left( D\left( y\right) y^{\prime \prime \prime
}+E\left( y\right) y^{\prime }y^{\prime \prime }+F\left( y\right) \left(
y^{\prime }\right) ^{3}\right) =0.  \label{eq.03}
\end{equation}

On the other hand, the Bennery-Gjevik equation takes the form of the
fourth-order ODE%
\begin{equation}
\left( A\left( y\right) -1\right) y^{\prime }+\varepsilon \left( B\left(
y\right) y^{\prime \prime }+C\left( y\right) \left( y^{\prime }\right)
^{2}\right) +\varepsilon ^{2}\left( \bar{D}\left( y\right) y^{\prime
}y^{\prime \prime \prime }+\bar{E}\left( y\right) y^{\left( 4\right)
}\right) =0.  \label{eq.04}
\end{equation}%
For more details on the application of symmetries on PDEs and on physical
problems we refer the reader in \cite{sj1,sj2,sj3,sj4,sj5,sj6,sj7} and
references therein.

We continue with the application of the singularity analysis to these ODEs.

\section{Integrability of the Benney equation}

\label{sec3}

We commence our discussion by studying the linearised Benney equation 
\begin{equation}
\left( A\left( y\right) -1\right) y^{\prime }+\varepsilon \left( B\left(
y\right) y^{\prime \prime }+C\left( y\right) \left( y^{\prime }\right)
^{2}\right) =0.  \label{eq.05}
\end{equation}

We substitute $y\left( \zeta \right) =y_{0}\left( \zeta -\zeta _{0}\right)
^{-\frac{1}{3}}$ into (\ref{eq.05}) which is reduced to the algebraic
equation%
\begin{eqnarray}
0 &=&\alpha _{1}\left( y_{0},R,a,\varepsilon \right) \left( \zeta -\zeta
_{0}\right) ^{8P-1}+a_{2}\left( y_{0},R,a,\varepsilon \right) \left( \zeta
-\zeta _{0}\right) ^{6P-1}+  \notag \\
&&+\alpha _{3}\left( y_{0},R,a,\varepsilon \right) \left( \zeta -\zeta
_{0}\right) ^{2P-2}+\alpha _{4}\left( y_{0},R,a,\varepsilon \right) \left(
\zeta -\zeta _{0}\right) ^{5P-2}.  \label{eq.06}
\end{eqnarray}%
From this we find that the leading-order behaviour $y\left( \zeta \right)
=y_{0}^{-\frac{1}{3}}\left( \zeta -\zeta _{0}\right), ~\ $in which $%
y_{0}=y_{0}\left( R,a\right), $ is given by a third-order polynomial
equation.

In order to find the resonances, i.e. the position of the other constant of
integration, we substitute 
\begin{equation}
y\left( \zeta \right) =y\left( \zeta -\zeta _{0}\right) ^{-\frac{1}{3}%
}+m\left( \zeta -\zeta _{0}\right) ^{-\frac{1}{3}+s}  \label{eq.07}
\end{equation}%
into (\ref{eq.05}) and we linearize around the parameter~$m$. There again
the leading-order terms provide the polynomial equation%
\begin{equation}
\left( S+1\right) \left( 3S+1\right) =0  \label{eq.08}
\end{equation}%
the zeros of which give the resonances; they are~$S=-1$ and $S=-\frac{1}{3}$%
. As we discussed above, the existence of the resonance, $S=-1$, is
important in order that the singularity exist and be movable. In particular
it is related with the constant of integration, $\zeta _{0}$. From the
second resonance we extract the information that the algebraic solution of
equation (\ref{eq.05}) is given by a Left Painlev\'{e} Series, with a step $%
\frac{1}{3}$, i.e. the algebraic solution is 
\begin{equation}
y\left( \zeta \right) =y_{0}\left( \zeta -\zeta _{0}\right) ^{-\frac{1}{3}%
}+y_{1}\left( \zeta -\zeta _{0}\right) ^{-\frac{2}{3}}+\dsum\limits_{I=2}^{%
\infty }y_{I}\left( \zeta -\zeta _{0}\right) ^{-\frac{1+I}{3}}  \label{eq.09}
\end{equation}%
in which $y_{1}$ is the second constant of integration and $%
y_{I}=y_{I}\left( y_{1},R,a\right) $.

Now in the case of the full Benney equation, when the nonlinear term $%
\varepsilon ^{2}$ is included, the situation is different because the new
terms contribute to the existence of the singularity and a substitution $%
y\left( \zeta \right) =y_{0}\left( \zeta -\zeta _{0}\right) ^{-\frac{1}{3}}$
into (\ref{eq.03}) fails to provide a singular leading-order behaviour. In
order to surpass this difficulty we perform the change of variables $y\left(
\zeta \right) =\frac{1}{Y\left( \zeta \right) }$ and in the new equation we
substitute $Y\left( \zeta \right) =Y_{0}\left( \zeta -\zeta _{0}\right) ^{P}$
from which we obtain the algebraic equation%
\begin{eqnarray}
0 &=&\beta _{1}\left( Y_{0},R,a,\varepsilon \right) \left( \zeta -\zeta
_{0}\right) ^{13P-1}+\beta _{2}\left( Y_{0},R,a,\varepsilon \right) \left(
\zeta -\zeta _{0}\right) ^{11P-1}+\beta _{3}\left( Y_{0},R,a,\varepsilon
\right) \left( \zeta -\zeta _{0}\right) ^{7P-2}+  \notag \\
&&+\beta _{4}\left( Y_{0},R,a,\varepsilon \right) \left( \zeta -\zeta
_{0}\right) ^{10P-2}+\beta _{5}\left( Y_{0},R,a,\varepsilon \right) \left(
\zeta -\zeta _{0}\right) ^{9P-3}+  \notag \\
&&+\beta _{6}\left( Y_{0},R,a,\varepsilon \right) \left( \zeta -\zeta
_{0}\right) ^{6P-3}+\beta _{7}\left( Y_{0},R,a,\varepsilon \right) \left(
\zeta -\zeta _{0}\right) ^{3P-1}.  \label{eq.10}
\end{eqnarray}

From this system we obtain the leading-order term 
\begin{equation}
Y\left( \zeta \right) =Y_{0}\left( \zeta -\zeta _{0}\right) ^{-\frac{1}{2}%
}~,~\left( Y_{0}\right) ^{4}=\frac{17\varepsilon ^{2}}{6c}  \label{eq.11}
\end{equation}%
with corresponding resonances 
\begin{equation}
s=-1~,~s=1~,~s=-\frac{17}{6}.  \label{eq.12}
\end{equation}%
Consequently the algebraic solution is expressed by a full Painlev\'{e}
Series with step $\frac{1}{6}$, that is%
\begin{equation}
Y\left( \zeta \right) =\dsum\limits_{J=-2}^{-\infty }Y_{J}\left( \zeta
-\zeta _{0}\right) ^{-\frac{3+J}{6}}+Y_{-1}\left( \zeta -\zeta _{0}\right)
^{-\frac{1}{2}-\frac{1}{6}}+Y_{0}\left( \zeta -\zeta _{0}\right) ^{-\frac{1}{%
2}}+Y_{1}\left( \zeta -\zeta _{0}\right) ^{-\frac{1}{2}+\frac{1}{6}%
}+\dsum\limits_{I=2}^{\infty }Y_{I}\left( \zeta -\zeta _{0}\right) ^{-\frac{%
3+I}{6}}.  \label{eq.13}
\end{equation}%
The latter Laurent expansion is replaced in equation (\ref{eq.03}) such that
to perform the consistency test. We find that expression (\ref{eq.13}) is a
solution for the Benney equation. 

At this point it is important to mention that the imaginary coefficient of
the leading-order term $Y_{0}$ is directly related with periodic solutions.
\ We continue our study with the Benney-Gjevik equation.

\section{Integrability of the Benney-Gjevik Equation}

\label{sec4}

For the Benney-Gjevik equation we apply the same procedure as for the full
Benney equation. Indeed, we replace $y\left( \zeta \right) =\frac{1}{Y\left(
\zeta \right) }$ and we perform the singularity analysis for the new
fourth-order ODE expressed in terms of $Y\left( \zeta \right) $.

We substitute $Y\left( \zeta \right) =Y_{0}\zeta ^{P}$ which gives us the
polynomial equation%
\begin{eqnarray}
0 &=&\gamma _{1}\left( Y_{0},R,a,\varepsilon ,W\right) \left( \zeta -\zeta
_{0}\right) ^{8P-1}+\gamma _{2}\left( Y_{0},R,a,\varepsilon ,W\right) \left(
\zeta -\zeta _{0}\right) ^{6P-1}+  \notag \\
&&+\gamma _{3}\left( Y_{0},R,a,\varepsilon ,W\right) \left( \zeta -\zeta
_{0}\right) ^{5P-2}+\gamma _{4}\left( Y_{0},R,a,\varepsilon ,W\right) \left(
\zeta -\zeta _{0}\right) ^{2P-2}+  \notag \\
&&+\gamma _{5}\left( Y_{0},R,a,\varepsilon ,W\right) \left( \zeta -\zeta
_{0}\right) ^{5P-4}  \label{eq.14}
\end{eqnarray}%
from which we extract the leading-order behaviour to be 
\begin{equation}
Y\left( \zeta \right) =Y_{0}\left( \zeta -\zeta _{0}\right) ^{\frac{2}{3}%
}~,~Y_{0}=Y_{0}\left( R,\varepsilon ,W\right) .  \label{eq.15}
\end{equation}%
in which~$Y_{0}$ is given by a fourth-order polynomial.

Moreover the resonances are calculated to be 
\begin{equation}
s=-1~,~s=\frac{8}{3}~,~s=\frac{10}{3}~,~s=\frac{17}{3}  \label{eq.16}
\end{equation}%
which indicates that the solution is expressed in terms of a Right Painlev%
\'{e} Series with step $\frac{1}{3}$, i.e.%
\begin{equation}
Y\left( \zeta \right) =Y_{0}\left( \zeta -\zeta _{0}\right) ^{\frac{2}{3}%
}+Y_{1}\left( \zeta -\zeta _{0}\right) +\dsum\limits_{I=2}^{\infty
}Y_{I}\left( \zeta -\zeta _{0}\right) ^{\frac{2+I}{3}}.  \label{eq.17}
\end{equation}

In addition, because the three resonances are positive, this indicates that
the leading-order behaviour is an attractor for the differential equation.
Finally, the Laurent expansion (\ref{eq.17}) satisfy the consistency test
which means that expression (\ref{eq.17}) is an analytic solution for the
Benney-Gjevik equation. The coefficients of (\ref{eq.17}) are ~$%
Y_{1-5}=0,~Y_{6}=-\frac{25}{18}Y_{6}\left( Y_{0}\left( R,\varepsilon
,W\right) \right) ~,~Y_{8}=arbitrary~,~Y_{9}=0$,~$Y_{9}=arbitrary$ etc. $\ $%
The integration constants are the coefficients $Y_{8},~Y_{10}$ and $Y_{17}$. 

\section{Conclusions}

\label{sec5}

In this work we studied the existence of travelling-wave solutions for the
Benney and the Benney-Gjevik equations by using the singularity analysis. In
particular we applied the invariants of the Lie symmetry vector $\partial
_{t}+c\partial _{x}$ and reduced the $1+1$ evolution equations (\ref{eq.01}%
), (\ref{eq.02}) to a third-order and a fourth-order ODE, respectively. We
proved that the these equations pass the Painlev\'{e} Test and consequently
are integrable.

For completeness in our analysis we have considered also the case where in
the Benney equation (\ref{eq.01}) parameter $\varepsilon ^{2}\rightarrow 0$
which is explicitly the case studied in \cite{ben1,ben2}. However, what is
worth mentioning is that all the leading-order terms can have complex
coefficients which means that the solutions can be periodic.

The importance of our analysis is that we were able to prove the
integrability for the Benney and the Benney-Gjevik equations while also to
present for the first time the algebraic expressions for the travelling-wave
solutions in terms of Laurent expansions.

We conclude this work by mentioning that the scopus of this work was to
prove the existence of analytic solutions for the evolution equations (\ref%
{eq.01}), (\ref{eq.02}) which in particular are singular perturbative
equations. Until now, according to our knowledge of the literature,
travelling-wave solutions have been found as approximate solutions at the
limits of the behaviour for these two equations by using perturbation
theory. In our analysis we considered the problem without assuming any
perturbative term and we were able to prove the existence of real solutions
(travelling-wave solutions) for arbitrary initial conditions, without
resorting to the Cauchy theorem for local existence.

\subsection*{Acknowledgements}

AP and GL were funded by Comisi\'{o}n Nacional de Investigaci\'{o}n Cient%
\'{\i}fica y Tecnol\'{o}gica (CONICYT) through FONDECYT Iniciaci\'{o}n
11180126. GL thanks to Department of Mathematics and to Vicerrector\'{\i}a
de Investigaci\'{o}n y Desarrollo Tecnol\'{o}gico at Universidad Cat\'{o}%
lica del Norte for financial support.

\appendix

\section{Coefficients for equations (\protect\ref{eq.01}) and (\protect\ref%
{eq.02})}

The coefficients of equation (\ref{eq.01}) are as follows \cite{benney66}%
\begin{eqnarray}
A\left( u\right) &=&2u^{2} \\
B\left( u\right) &=&-\frac{8}{15}Ru^{6}+\frac{2}{3}u^{3}\cot a \\
C\left( u\right) &=&-\frac{16}{5}Ru^{5}+2u^{2}\cot a \\
D\left( u\right) &=&-2u^{4}-\frac{32}{68}R^{2}u^{10}+\frac{40}{63}Ru^{7}\cot
a \\
E\left( u\right) &=&-\frac{52}{3}u^{3}+\frac{433}{63}R^{2}u^{9}+\frac{392}{45%
}Ru^{6}\cot a \\
F\left( u\right) &=&-14u^{2}-29R^{2}u^{8}+\frac{64}{5}Ru^{5}\cot a,
\end{eqnarray}%
where $R$ denotes the Rayleigh's number, $a$ is the inclination of the plane
on which the motion occurs, while the dependent variable $u\left( t,x\right) 
$ describes the surface of the flow. \ While for equation (\ref{eq.02})
coefficients $\bar{D}\left( u\right) $ and $\bar{E}\left( u\right) $ are 
\begin{equation}
\bar{D}\left( u\right) =3Wu^{2}~,~\bar{E}\left( u\right) =3Wu^{3}~
\end{equation}%
in which $W$ is a parameter of order $a^{-2}$ \cite{gjevik}.

\end{document}